\documentclass[prd,aps,twocolumn,nofootinbib,floatfix,longbibliography, superscriptaddress]{revtex4-1}
\usepackage{graphicx}
\usepackage{bm}
\usepackage{mathtools}
\usepackage{hyperref}
\usepackage{amsmath}
\usepackage{mathrsfs}
\usepackage{color}
\usepackage{orcidlink}

\usepackage{hyperref}
\hypersetup{
 pdftitle={Interpolated kilonova spectra models: necessity for a phenomenological, blue component in the fitting of AT2017gfo spectra}, %
 pdfauthor={Ristic}, %
 colorlinks=true,
 citecolor={blue},
}

\newcommand\edited[1]{{\color{black}#1}}

\usepackage{soul}

\def\RIT{Center for Computational Relativity and Gravitation, Rochester Institute of Technology, Rochester, New York
 14623, USA}
\def\XCP{Computational Physics Division, Los Alamos National Laboratory, Los Alamos, NM, 87545, USA}
\def\CTA{Center for Theoretical Astrophysics, Los Alamos National Laboratory, Los Alamos, NM 87545, USA}
\def\CCSS{Computer, Computational, and Statistical Sciences Division, Los Alamos National Laboratory, Los Alamos, NM
 87545, USA}
\def\TD{Theoretical Division, Los Alamos National Laboratory, Los Alamos, NM 87545, USA}

\def\UA{The University of Arizona, Tucson, AZ 85721, USA}
\def\NM{Department of Physics and Astronomy, The University of New Mexico, Albuquerque, NM 87131, USA}

\def\GWU{The George Washington University, Washington, DC 20052, USA}

\graphicspath{{figures/}}

\begin{document}

\title{Interpolated kilonova spectra models: necessity for a phenomenological, blue component in the fitting of AT2017gfo spectra}

\author{Marko Risti\'c\,\orcidlink{0000-0001-7042-4472}}
\affiliation{\RIT}
\author{Richard O'Shaughnessy\,\orcidlink{0000-0001-5832-8517}}
\affiliation{\RIT}
\author{V. Ashley Villar\,\orcidlink{0000-0002-5814-4061}}
\affiliation{Department of Astronomy \& Astrophysics, The Pennsylvania State University, University Park, PA 16802, USA}
\affiliation{Institute for Computational \& Data Sciences, The Pennsylvania State University, University Park, PA 16802, USA}
\affiliation{Institute for Gravitation and the Cosmos, The Pennsylvania State University, University Park, PA 16802, USA}
\author{Ryan~T. Wollaeger\,\orcidlink{0000-0003-3265-4079}}
\affiliation{\CCSS}
\affiliation{\CTA}
\author{Oleg Korobkin\,\orcidlink{0000-0003-4156-5342}}
\affiliation{\CTA}
\affiliation{\TD}
\author{Chris~L. Fryer\,\orcidlink{0000-0003-2624-0056}}
\affiliation{\CTA}
\affiliation{\CCSS}
\affiliation{\UA}
\affiliation{\NM}
\affiliation{\GWU}
\author{Christopher~J. Fontes\,\orcidlink{0000-0003-1087-2964}}
\affiliation{\CTA}
\affiliation{\XCP}
\author{Atul Kedia\,\orcidlink{0000-0002-3023-0371}}
\affiliation{\RIT}

\date{\today}

\begin{abstract}
In this work, we present a simple interpolation methodology for spectroscopic time series,
based on conventional interpolation techniques (random forests) implemented in widely-available libraries. 
We demonstrate that our existing library of simulations is sufficient for training, \edited{producing interpolated}
spectra \edited{that} respond sensitively to varied ejecta parameter, post-merger time, and viewing angle inputs.  We
compare our interpolated spectra to the AT2017gfo spectral data, and find parameters similar to our previous
inferences using broadband light curves.
However, the spectral observations have significant systematic short-wavelength residuals relative to our models, which we
cannot explain within our existing framework. Similar to previous studies,  we argue \edited{that} an additional blue component is required. \edited{We consider a radioactive heating source as a third component characterized by} light, slow-moving, lanthanide-free ejecta with $M_{\rm th} = 0.003~M_\odot$, $v_{\rm th} = 0.05$c, and $\kappa_{\rm th} = 1$ cm$^2$/g. When \edited{included as part of our radiative transfer simulations, our choice of third component reprocesses blue photons into lower energies, having the opposite effect and further accentuating the blue-underluminosity disparity in our simulations}.  \edited{As such, we are unable to overcome short-wavelength deficits at later times using an additional radioactive heating component}, indicating the need for a more sophisticated modeling treatment.
\end{abstract}

\maketitle

\section{Introduction}
\label{sec:intro}

The detection of the joint gravitational- and electromagnetic-wave emission from binary neutron star
merger GW170817 \citep{LIGO-GW170817-mma} and its electromagnetic counterpart AT2017gfo \citep{Tanvir_2017} has
initiated an era of precision kilonova observations. Several studies interpreted the observations of AT2017gfo shortly
after detection by comparing to simple kilonova models \citep{Villar_2017, 2017ApJ...848L..17C, 2017ApJ...850L..37P}
consisting of one or more groups of homologously-expanding material. 
Motivated both by binary merger simulations and the inability to fit observations with one component, at least two
components are customarily employed, with properties loosely associated with two expected features of merger
simulations: promptly ejected material (the ``dynamical'' ejecta), associated with tidal tails or shocked
material at contact; and material driven out on longer timescales by properties of the remnant system (the ``wind'' ejecta) \citep{2019ARNPS..69...41S}.
However, many of these simple kilonova models lack important physical features expected from neutron star merger
simulations, including full radiative transfer and opacities, as well as anisotropic outflow and emission.  
More recent modeling efforts increasingly incorporate these features, including sophisticated treatments of relevant
kilonova microphysics \citep{2021arXiv210101201B, 2022MNRAS.512.1499R, 2022ApJ...933...22K, 2023MNRAS.520.2558B}.  
Due to the high simulation cost,  many groups have resorted to surrogate models for the kilonova
outflow,  to reduce the computational cost associated with inference with these more complex models \citep{gwastro-mergers-em-CoughlinGPKilonova-2020, 2021arXiv211215470A, Ristic22, 2022MNRAS.516.1137L}. 

Despite the increasingly sophisticated models being brought to bear to interpret AT2017gfo, the shorter-wavelength
$g$-band flux that was observed in AT2017gfo cannot be easily described using only a conventional two-component model \citep{2018ApJ...855L..23A, Ristic22, Kawaguchi_2020, 2022ApJ...933...22K}.   While a ``third component" could  resolve  this underluminosity, as yet many physical processes are being investigated to drive such an outflow and thereby specify how its properties relate to other system parameters, including  ejecta shock breakout \citep{Nicholl21} and central engine sources \citep{2013ApJ...776L..40Y, 2014MNRAS.439.3916M, 2018ApJ...861L..12L, 2018ApJ...861...55M, 2019MNRAS.483.1912P, 2019ApJ...880...22W, 2021ApJ...915L..11A}.   Of course, this underluminosity could also in part reflect insufficiently well-understood kilonova systematics; see, e.g., \cite{2019Natur.574..497W, Kawaguchi_2020, 2021ApJ...918...44B, 2021ApJ...906...94Z}.

Most interpretations of \edited{kilonova observations} have relied on broadband photometry, in part owing to the relative sparsity of available spectra for AT2017gfo (and other kilonovae).  Fast interpolated  models for (anisotropic) kilonova spectra, computed with state of the art opacities, could provide a new avenue to resolve key uncertainties about AT2017gfo and other kilonovae.  Several recent projects have demonstrated the high potential return of comparing AT2017gfo to kilonova spectral models \citep{2021MNRAS.506.3560G, 2023Natur.614..436S}.  
In this work, we present a detailed interpolation scheme for kilonova spectra which allows for continuous spectral modeling across time and viewing angle. We showcase our ability to produce interpolated spectra outputs at various ejecta parameters, times, and angles. In accordance with previous studies, we identify the need for a third component in order to partially match our model's $g$-band spectral energy density to that of AT2017gfo.
Our method can be easily applied to any modestly-sized archive of adaptively-learned \edited{astrophysical transient spectra} simulations.

The paper is organized as follows. Section~\ref{sec:intp} discusses our simulation training library and associated spectra interpolation methodology. In Section~\ref{sec:2c}, we compare our interpolated spectra to those observed for the kilonova AT2017gfo and present the best-fitting ejecta parameters that reproduce the AT2017gfo spectra assuming a two-component model. In Section~\ref{sec:3c}, we explore the effects of including a third, low-opacity component to supplement shorter wavelength ($g$-band) flux in our simulations. We summarize our findings in Section~\ref{sec:conclusion}.

\section{Interpolation Methodology}
\label{sec:intp}

\subsection{Simulation Description}
\label{sec:sim_setup}

Unless noted otherwise, we consider a two-component kilonova model with a lanthanide-rich equatorial dynamical ejecta component and a lanthanide-poor axial wind ejecta component as described in \cite{kilonova-lanl-WollaegerNewGrid2020, 2021ApJ...910..116K} and motivated by numerical simulations \citep{2019ARNPS..69...41S,just23}. Each component is parameterized by a mass and velocity such that $M_{\rm{d}}$, $v_{\rm{d}}$ and $M_{\rm{w}}$, $v_{\rm{w}}$ describe the dynamical and wind components' masses and velocities, respectively. The morphology for the dynamical component is an equatorially-centered torus, whereas the wind component is represented by an axially-centered peanut component; Figure~1 of \cite{kilonova-lanl-WollaegerNewGrid2020} displays the \edited{torus-peanut}, or ``TP," schematic corresponding to the  morphologies employed in this work \citep[see][for detailed definition]{2021ApJ...910..116K}. The lanthanide-rich dynamical ejecta is a result of the $r$-process nucleosynthesis from a neutron-rich material with a low electron fraction ($Y_{\rm{e}} \equiv n_{\rm{p}}/(n_{\rm{p}} + n_{\rm{n}})$) of $Y_e = 0.04$ with elements reaching the third $r$-process peak ($A \sim 195$), while the wind ejecta originates from higher $Y_e = 0.27$ which encapsulates elements between the first ($A \sim 80$) and second ($A \sim 130$) $r$-process peaks. The detailed breakdown of the elements in each component can be found in Table~2 of \cite{kilonova-lanl-WollaegerNewGrid2020}.

We use \texttt{SuperNu}, a Monte Carlo code for simulation of time-dependent radiation transport with matter in local thermodynamic equilibrium, to create simulated kilonova spectra $F_{\lambda, \rm sim}$ assuming the aforementioned two-component model \citep{SuperNu}. Both components are assumed to have fixed composition and morphology for the duration of each simulation. \texttt{SuperNu} uses radioactive power sources calculated from decaying the $r$-process composition from the \texttt{WinNet} nuclear reaction network \citep{2012ApJ...750L..22W,Korobkin_2012}. These radioactive heating contributions are also weighted by thermalization efficiencies introduced in \cite{Barnes_2016} \citep[see][for a detailed description of the adopted nuclear heating]{Wollaeger2018}. We use detailed opacity calculations via the tabulated, binned opacities generated with the Los Alamos suite of atomic physics codes \citep{2015JPhB...48n4014F,2020MNRAS.493.4143F,nist_lanl_opacity_database}. Our tabulated, binned opacities are not calculated for all elements; therefore, we produce opacities for representative proxy elements by combining pure-element opacities of nuclei with similar atomic properties \citep{2020MNRAS.493.4143F}. Specifics of the representative elements for our composition are given in \cite{kilonova-lanl-WollaegerNewGrid2020}.

The \texttt{SuperNu} outputs are anisotropic simulated spectra $F_{\lambda, \rm sim}$, post-processed to a source distance of $10$ pc, in units of erg s$^{-1}$ cm$^{-2}$ \AA$^{-1}$. The spectra are binned into 1024 equally log-spaced wavelength bins spanning $0.1 \leq \lambda \leq 12.8$~microns. For the purposes of this work, we consider the spectral evolution across 60 equally log-spaced times between 0.125 and 20.75 days post-merger. However, many of the spectra in our training library extend out to even later times. As we only consider anisotropic simulations in this study, we extract simulated spectra using 54 angular bins, uniformly spaced as $-1 \leq \cos \theta \leq 1$ for the angle $\theta$ between the line of sight and the symmetry axis.

\subsection{Training Set Generation}
\label{sec:trainingset}

\edited{The follow description describes the approach taken to generate the simulation library in \cite{Ristic22}}. Our training library of $412$ kilonova spectra and light-curve simulations was constructed using iterative simulation placement guided by Gaussian process variance minimization. In previous work, we focused solely on light-curve interpolation; as such, new simulations were placed with parameter combinations \edited{that} were identified as having the largest bolometric luminosity variance by our Gaussian process regression approach. In other words, we placed new simulations in regions of parameter space where our bolometric luminosity interpolation root-mean-square uncertainty was largest. Equation~\ref{eq:rms} shows the Gaussian process variance $s(\vec{x})^2$

\begin{equation}
s(\vec{x})^2 = k(\vec{x}, \vec{x}) - k(\vec{x}, \vec{x}_a)k(\vec{x}_a, \vec{x}_{a'})^{-1}_{aa'}k(\vec{x}_{a'}, \vec{x})
\label{eq:rms}
\end{equation}
where $\vec{x}$ is the vector of input parameters, $\vec{x}_a$ is the training data vector, $s(\vec{x})^2$ is the variance of the Gaussian process prediction, the function $k(\vec{x}, \vec{x}')$ is the kernel of the Gaussian process, and the indices $a, a'$ are used to calculate the covariance between inputs $\vec{x}$ and training data $\vec{x}_{a}, \vec{x}_{a'}$ such that if $a = a'$, the variance is 0.

In the context of this work, the only relevance of the aforementioned light curves is to \edited{explain the process of constructing} the \edited{original} simulation library. The spectra used in this work have the same parameters as the light curves used for our light-curve interpolation approach in \cite{Ristic22}. \edited{No additional simulations were produced for the purposes of this work; all training data came from the simulation library presented in \cite{Ristic22}.}

\edited{The original training data library consists of} $412$ total simulations calculated at $60$ times ($54$ angles) each \edited{for a total of} $24720$ (22248) spectra evaluated at $1024$ wavelength bins. Due to the sheer volume of data in our training set, we do not perform any coordinate transformations, but rather interpolate directly in our ejecta parameter space and time or angle. \edited{However, the large data volume incurs a high computational cost, most notably high memory usage during training. For the remainder of the work, unless otherwise noted, we downsample our data to only include spectra evaluated between $1.4$ and $10.4$ days for wavelengths above $0.39$ microns (the lower limit of the $g$-band) and below $2.39$~microns (the upper limit of the $K$-band). Downsampling reduces the dataset to $412$ total simulations calculated at $24$  times for a total of $9888$ spectra evaluated at $384$ wavelength bins. The angular bins can be similarly downsampled from $54$ to $27$ to get a comparable data volume.} For simplicity, all subsequent discussion will refer to interpolation in time; however, all instances of time as an interpolation parameter can be directly replaced with angle.

\subsection{Spectrum Interpolation Approach}
\label{sec:rf}

Our spectrum simulation setup and interpolation scheme presented in this work differ slightly from the approach described in Section~\ref{sec:trainingset}. As before, our inputs are the four ejecta parameters describing our two-component kilonova model, with the addition of post-merger time in days, such that we have a five-dimensional input $\vec{x} = (M_d, v_d, M_w, v_w, t)$. \edited{For completeness, the angle $\theta$ can remain unfixed, allowing a six-dimensional input $\vec{x} = (M_d, v_d, M_w, v_w, t, \theta)$ at greater computational cost.}
For each fixed viewing angle, our interpolation output is the spectral energy density $F_\lambda$  associated with that viewing angle in units of erg s$^{-1}$ cm$^{-2}$ \AA$^{-1}$. We favor a random forest interpolation scheme due to its enhanced recovery of detailed spectral features compared to the Gaussian process approach. This choice comes at the cost of losing an inherent uncertainty prediction \edited{that} is associated with Gaussian process interpolation output. \edited{We recognize the existence of random forest uncertainty calculation modules, but have been unable to successfully incorporate them in our study}.

Random forests are a sub-class of grouped decision-tree structures \edited{that} can be used for regression applications. The following summary is adapted from the \texttt{scikit-learn} documentation on decision trees \citep{sklearn}\footnote{\url{https://scikit-learn.org/stable/modules/tree.html}}. An individual tree in a random forest recursively partitions the spectral flux density samples $F_\lambda$ for a set of five-dimensional input parameters $\vec{x}$ from the training set via a series of decisions, commonly referred to as branches, based on a randomly-selected threshold value. This threshold value $t_i$ can be thought of as a piecewise function \edited{that} divides the samples into two groups, or leaf nodes, $Q_i$: one where all of the samples meet the branch threshold, $Q_i^{left}$, and another where none of the samples meet the branch threshold, $Q_i^{right}$,
\begin{align}
Q_i^{left} &= \{ F_\lambda \ | \ F_\lambda \leq t_i \} \\
Q_i^{right} &= Q_i \ \backslash \ Q_i^{left}.
\end{align}
These thresholds are generated recursively, with each subsequent leaf node $Q_i^{left/right}$ being re-partitioned until a specified recursion termination step is reached. The tree is then left with a total of $m$ leaf nodes, each of which contains $n_m$ spectral flux density values $F_\lambda$ from the original dataset. The predicted spectral flux density for each leaf node is given by 
\begin{equation}
\overline{F}_{\lambda,m} = \frac{1}{n_m} \sum_{F_\lambda \in Q_m} F_\lambda
\end{equation}
with an associated likelihood for each node defined by a mean-squared error
\begin{equation}
\mathcal{L}(Q_m) = \frac{1}{n_m} \sum_{F_\lambda \in Q_m} (F_\lambda - \overline{F}_{\lambda, m})^2 \,,
\end{equation}
where $m$ represents the given random forest node, $\overline{F}_{\lambda,m}$ is the learned mean value for node $m$, $n_m$ is the number of samples in node $m$, $Q_m$ is the training data in node $m$, and $\mathcal{L}(Q_m)$ is the probability of the learned mean value $\overline{F}_{\lambda,m}$ given partitioned training data $Q_m$. The learned mean value predictions in each node are weighted by their nodes' likelihoods to produce an individual tree's prediction for a given input $\vec{x}$. The random forest considers the outputs of all decision trees and uses majority voting to create the final interpolation prediction $F_{\lambda, \rm intp}$ for each angular bin.
Using the independent  random-forest estimates for each angular bin, we can interpolate, as needed, these predictions versus viewing angle, reconstructing a continuous estimate for  the flux as a function of simulation parameters, time, and viewing angle.
Conversely, we can repeat the procedure described above, exchanging time and angle, to produce a random-forest interpolation versus simulation parameters and angle, which we can interpolate in time as needed.

\begin{figure}
\includegraphics[trim={0 0 0 0}, clip, width=\columnwidth]{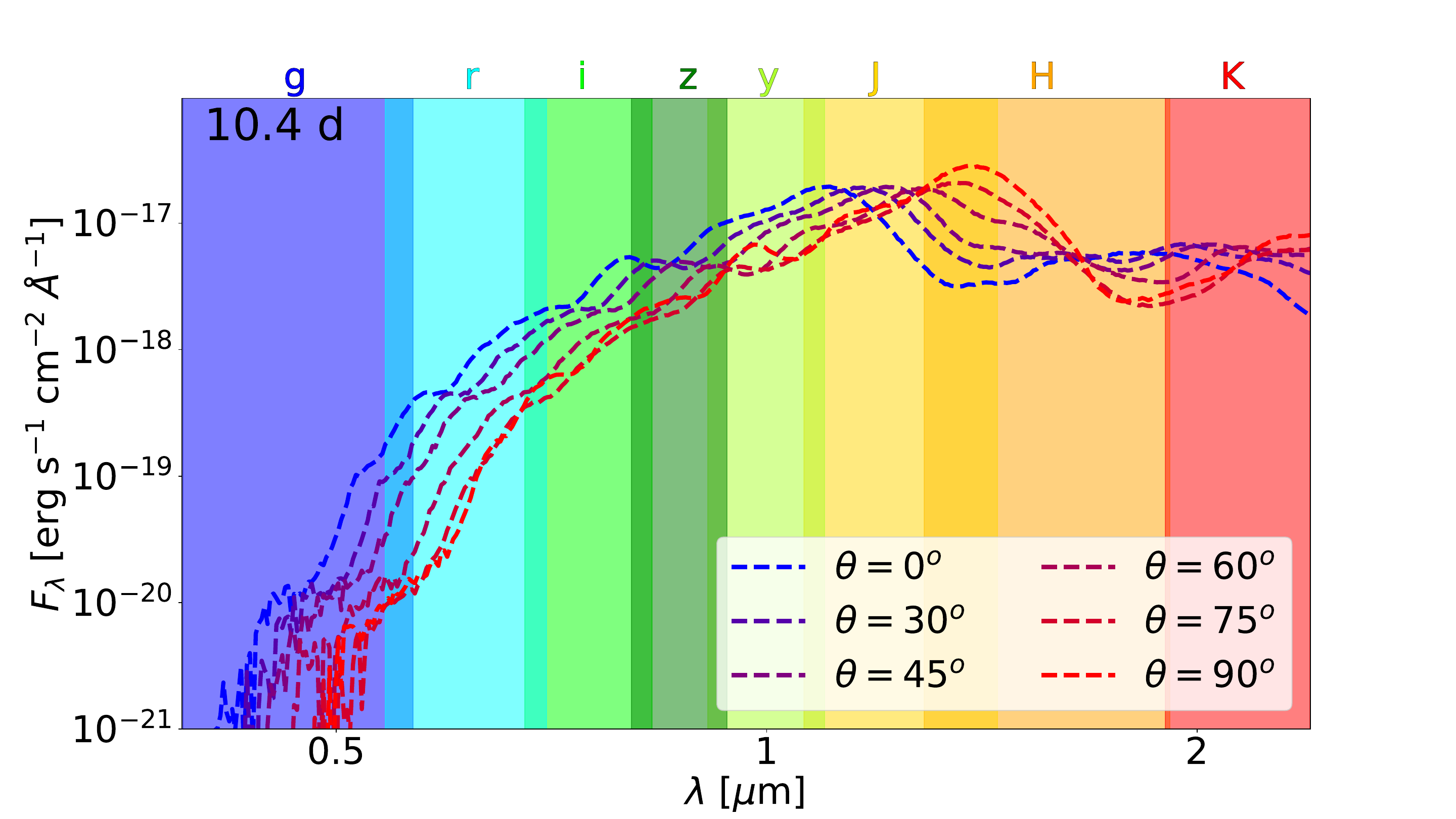}
\includegraphics[trim={0 0 0 0}, clip, width=\columnwidth]{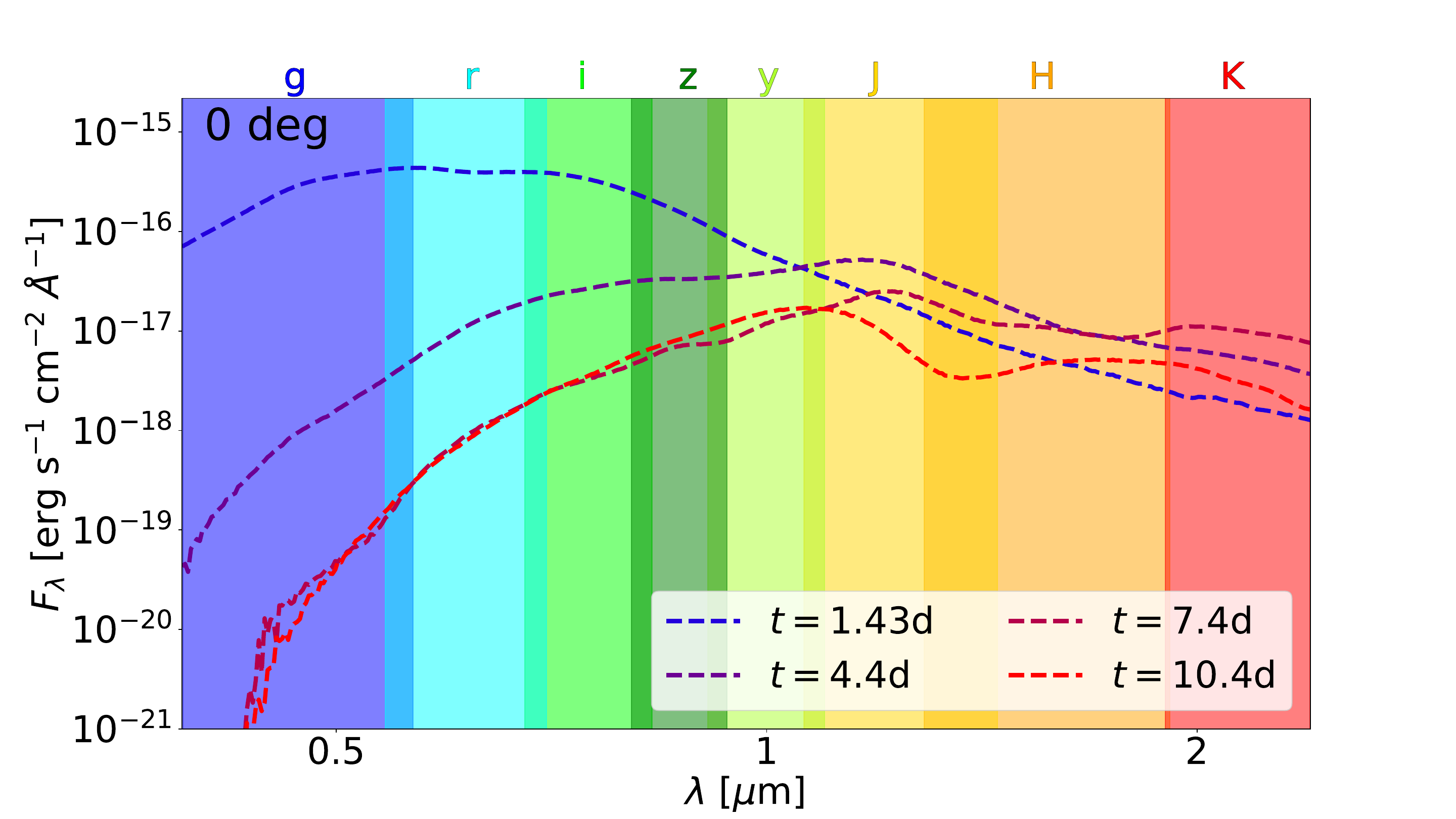}
\caption{Off-sample interpolated spectra at different viewing angles at a fixed time of 10.4 days (top) and different times at a fixed viewing angle of 0~degrees (bottom) with the same ejecta parameters as in Figure~\ref{fig:off_sample}. The spectra in the top figure exhibit the characteristic lanthanide-curtaining effect at shorter wavelengths as the dynamical ejecta becomes dominant at larger angles. The spectra in the bottom figure show the expected shift toward brighter spectral energy density in infrared wavelengths at later times.}
\label{fig:variation}
\end{figure}

\begin{figure}
\centering
\includegraphics[trim={0 0 0 2.1cm}, clip, width=\columnwidth]{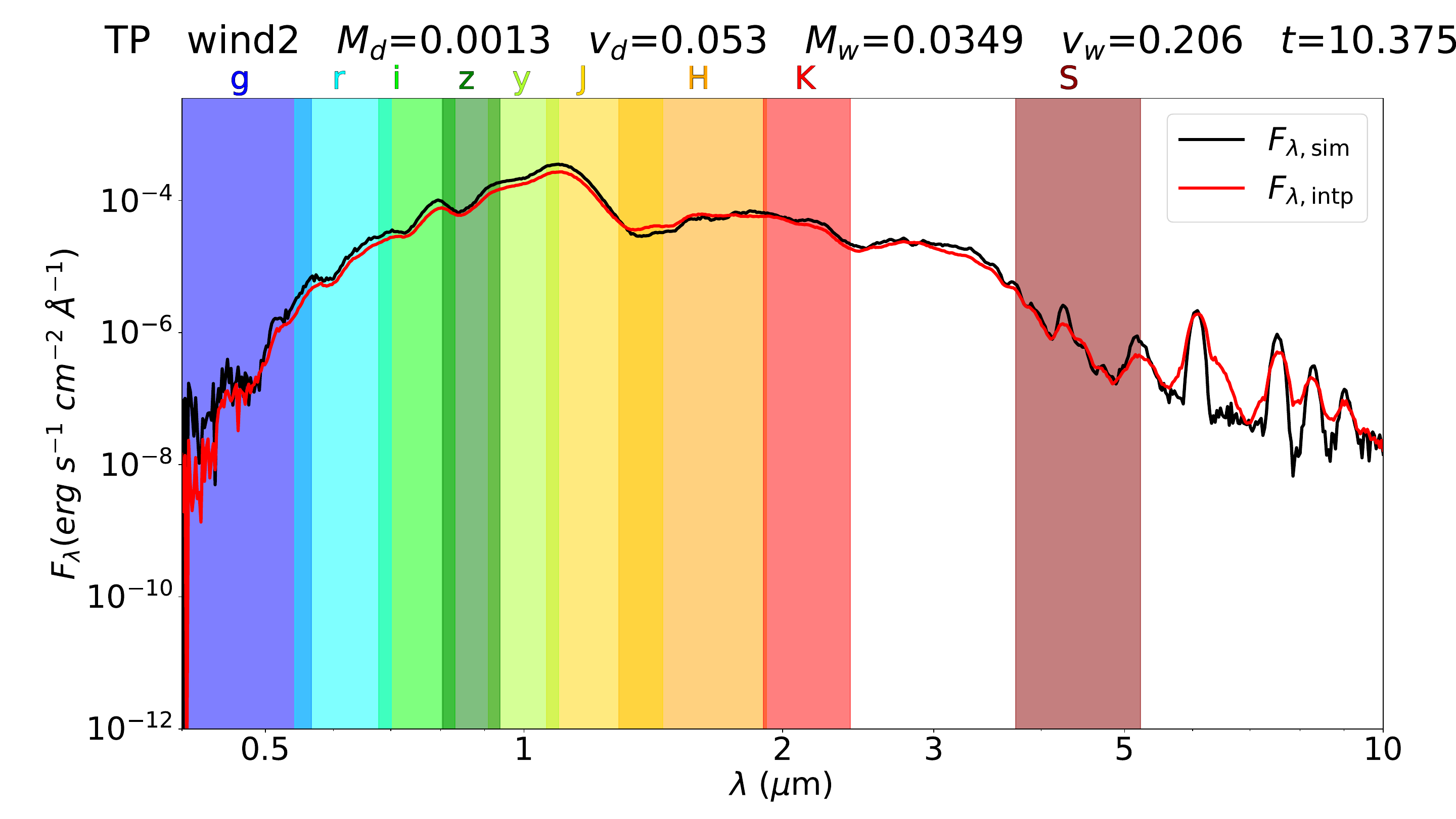}
\includegraphics[trim={0 0 0 2.1cm}, clip, width=\columnwidth]
{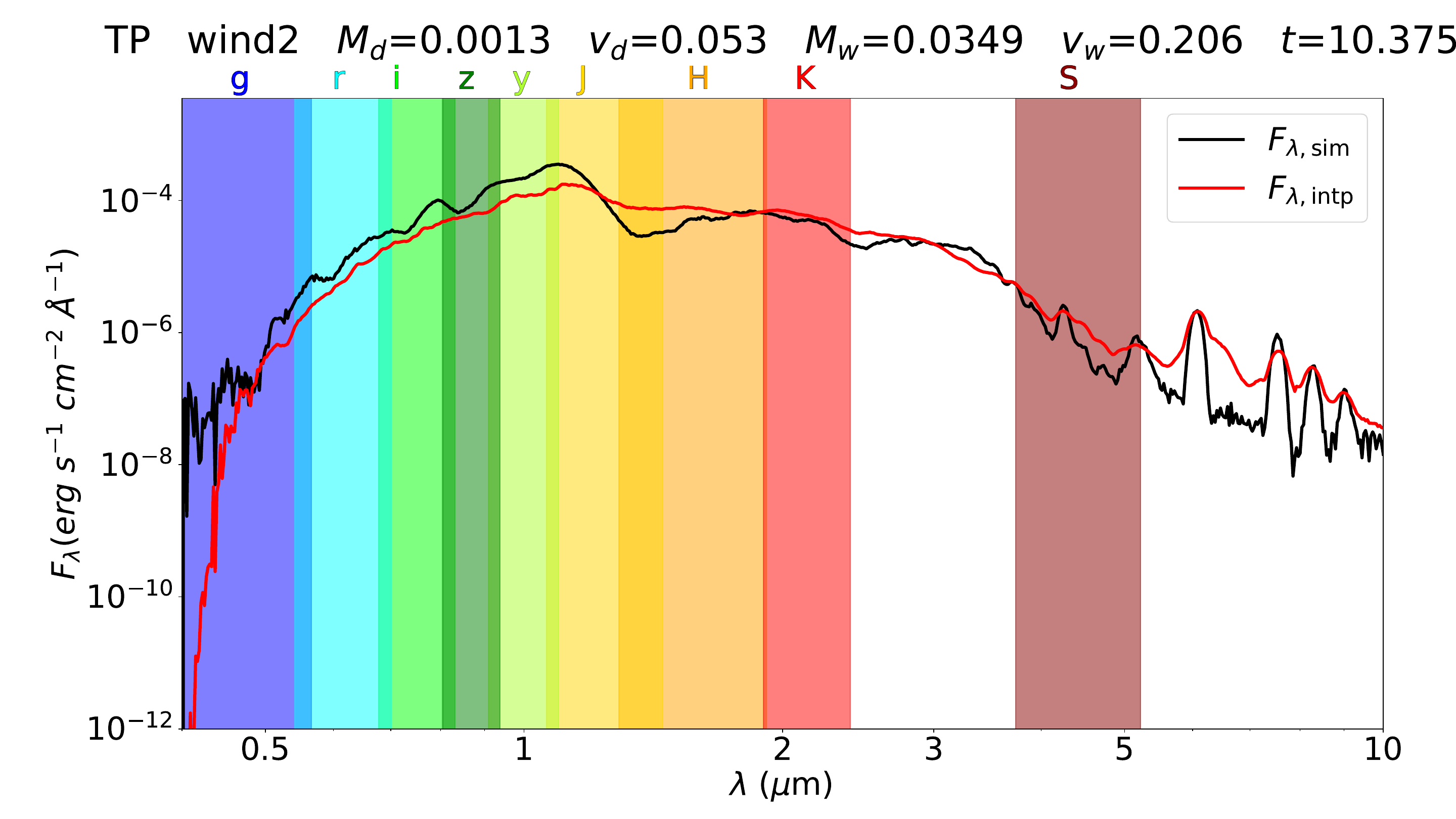}
\caption{\edited{Off-sample comparison of simulation data, in black, compared to an interpolated spectrum generated using the simulation input parameters, in red}. The simulation was evaluated for input parameters $M_{\rm{d}} = 0.0013$, $v_{\rm{d}} = 0.053$, $M_{\rm{w}} = 0.0349$, $v_{\rm{w}} = 0.206$, and $t = 10.4$ assuming a fixed viewing angle bin $\theta \leq {\sim}16^\circ$ and source distance of $10$~pc. Masses, velocities, time, and angle are in units of $M_{\odot}$, $c$, days, and degrees, respectively. \edited{Top: The off-sample prediction, in red, from a random forest interpolator trained without hyperparameter constraints and significantly higher computational resource cost. The unbounded computational cost allows for particularly accurate feature recovery, especially at wavelengths past 5~microns. Bottom: Same as above, except with hyperparameter constraints resulting in a much more computationally inexpensive model. The model prediction is noticeably smoother, however it still captures the general profile of the spectrum and the tops of the peak features past 5~microns.}}
\label{fig:off_sample}
\end{figure}

As previously mentioned, time and angle can be interchangeably included as interpolation parameters in our framework. Figure~\ref{fig:variation} showcases examples of using one of these parameters as an interpolation parameter and keeping the other fixed. The ejecta parameters in both panels were fixed to match those in Figure~\ref{fig:off_sample}; as such, all variations in Figure~\ref{fig:variation} are due \emph{solely to the fixed parameter}, $\theta$ or $t$, displayed in the figure legend. For convenience, we also overplot colored wavelength regions corresponding to the LSST \textit{grizy}, 2MASS \textit{JHK}, and the Spitzer 4.5 micron ``\textit{S}" broadband filters.

The top panel displays spectra at a fixed time of $10.4$ days and the changes in spectral features as the viewing angle is increased from $0$ (axial) to $90$ degrees (equatorial). \edited{In general, $F_\lambda$ tends to decrease as the viewing angle increases, moving away from the jet axis toward the plane in which the accretion disk lies. This behavior is expected as our low $Y_e$ dynamical ejecta component, concentrated in a torus near the plane, synthesizes heavier elements that contribute to higher opacity as $\theta$ increases, commonly referred to as lanthanide curtaining.}

The bottom panel, in a similar fashion, indicates how the spectra at a fixed viewing angle of $0$ degrees evolve over time between $1.43$ and $10.4$ days. \edited{The flux at the earliest times peaks in the lower-wavelengths bands before the system has had a chance to lose energy via expansion and thermal emission. At later times, as the system cools, the peak flux migrates to redder wavelengths and in some cases distinct spectral features begin to form.}

Figure~\ref{fig:off_sample} compares the predictions of our interpolation technique to a single out-of-sample simulation, \edited{evaluated at all simulation wavelengths} at a specific time and viewing angle.
The random forest prediction agrees remarkably with the underlying simulation data. \edited{The full wavelength range was considered in this instance due to the sharp, pronounced features past $\lambda > 5$~microns. The panels of Figure~\ref{fig:off_sample} show the same off-sample prediction using a more (less) computationally expensive approach during training in the top (bottom) panel}. 

Our spectra interpolation tool, as well as sample use cases, can be found at \url{https://github.com/markoris/rf_spec_intp}.

\section{Two-Component Analysis}
\label{sec:2c}
\subsection{AT2017gfo Observational Dataset}
\label{sec:obs}

In addition to serving as an interpolation training set, our simulated spectra can also inform us about which model parameters recreate the observed spectra for AT2017gfo. We use an observational dataset consisting of the ten X-shooter spectra originally published in \cite{2017Natur.551...67P} and \cite{2017Natur.551...75S}, which have been re-reduced and recalibrated by the ENGRAVE collaboration \citep{2020A&A...643A.113A}. The details of the spectral data cleaning, including an additional flux calibration step, are described in \cite{2022MNRAS.515..631G}. 
Throughout this work, unless specified otherwise, we use the flux-corrected, smoothed, joined spectra $F_{\lambda,\rm obs}$ obtained from the ENGRAVE data release\footnote{\url{http://www.engrave-eso.org/AT2017gfo-Data-Release}}. The data span a wavelength range of roughly 0.33 to 2.4~microns, with a couple \edited{of} spectra having a slightly shorter wavelength range.

\subsection{Fitting SuperNu Simulations to AT2017gfo}
\label{sec:2c_fit}

As described in Section~\ref{sec:sim_setup}, \texttt{SuperNu} outputs kilonova spectra $F_{\lambda,\rm{sim}}$ at a distance of $10$~pc across 1024 log-spaced wavelength bins $\lambda_k \text{ for } k = 0, 1, ..., 1023$ between 0.1 and 12.8~microns. The subscript $k$ notation hereafter refers to these 1024 \texttt{SuperNu} wavelength bins. For comparison between simulated and observed data, we scale the simulated spectra to a distance of 40~Mpc to match the distance at which AT2017gfo was observed. \edited{We fix the viewing angle to the first simulation angular bin ($\theta \leq {\sim}16^\circ$)}. 

We also downsample the observational data $F_{\lambda, \rm obs}$ such that each new observational wavelength bin corresponds to a \texttt{SuperNu} wavelength bin $\lambda_k$ and contains a new observational flux value $\hat{F}_{\lambda, \text{obs}, k}$ defined as

\begin{equation}
\hat{F}_{\lambda, \text{obs}, k} = \frac{1}{N_k}\sum_i F_{\lambda, \text{obs}, i} \text{ for } \lambda_k \leq \lambda_i < \lambda_{k+1},
\end{equation}
where $N_k$ is the number of original observational wavelength data points $\lambda_i$ \edited{that} are downsampled into the relevant \texttt{SuperNu} wavelength bin $\lambda_k$. From this point on, we refer to the rebinned, downsampled observational data as $\hat{F}_{\lambda, \rm obs}$. Due to the difference in wavelength ranges between our observed and simulated data sets, we are only able to compare the observed data to \emph{at most} 361 \texttt{SuperNu} wavelength bins between $0.33$ and $2.4$~microns. Our only other observational data processing involves removing portions of the observed spectra that exhibit telluric effects or artifacts from the stitching process. The gaps corresponding to the removed data are located around 0.6, 1, 1.4, and 1.9~microns. \edited{The data preprocessing described here is independent of the data-volume reduction steps described in Section~\ref{sec:trainingset}.}

\begin{figure}
\centering
\includegraphics[page=1, width=.95\columnwidth]{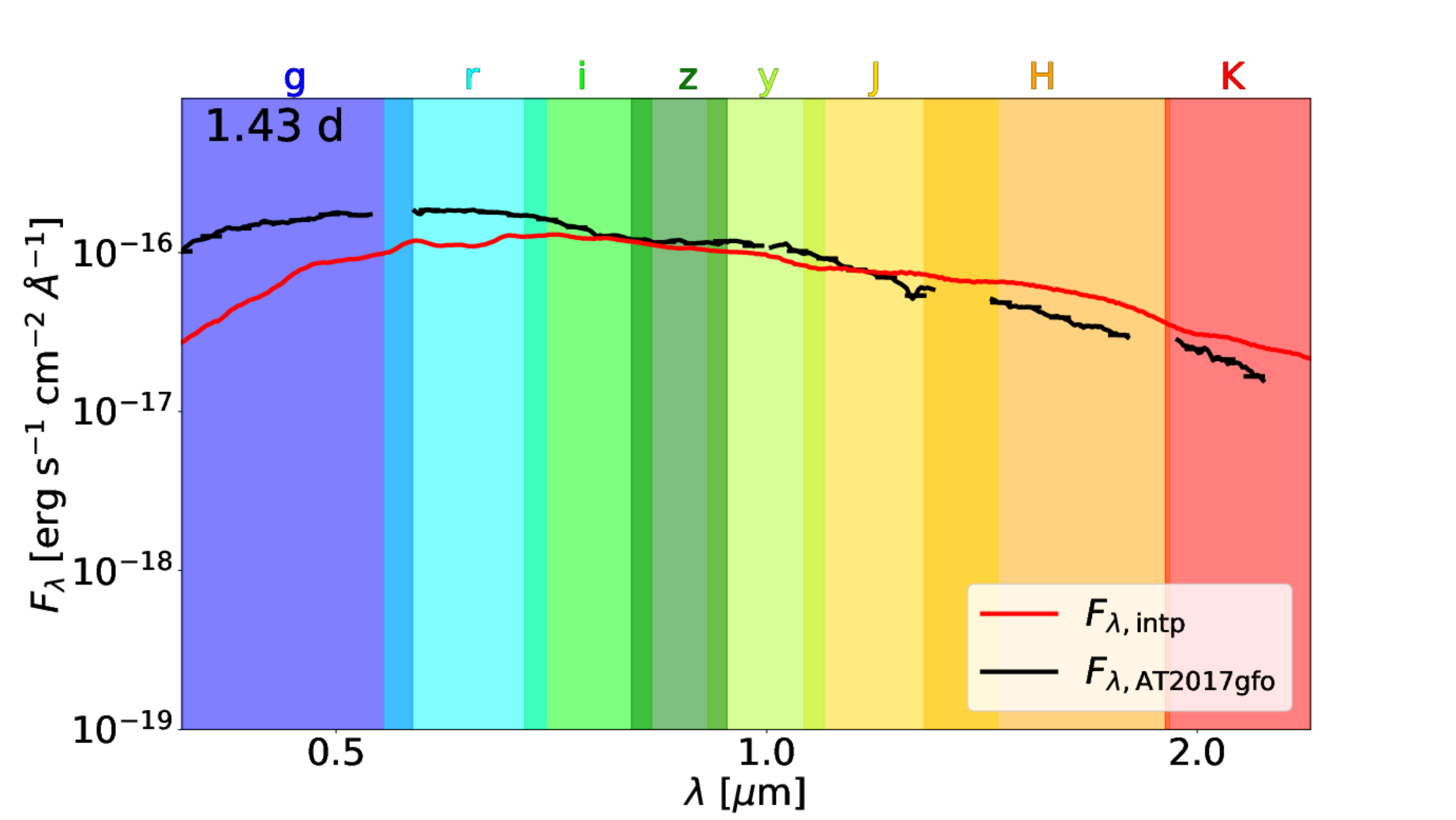}
\includegraphics[page=4, width=.95\columnwidth]{spec_intp_fits_optim_trimdata_obserror.pdf}
\includegraphics[page=7, width=.95\columnwidth]{spec_intp_fits_optim_trimdata_obserror.pdf}
\includegraphics[page=10, width=.95\columnwidth]{spec_intp_fits_optim_trimdata_obserror.pdf}
\caption{Interpolated, two-component kilonova spectra fitted to AT2017gfo observed spectra at $t = 1.43$ (top), $t = 4.4$ (upper middle), $t = 7.4$ (lower middle), and $t = 10.4$ (bottom) days. Each fit was calculated using Equation~\ref{eq:chi2} by only considering spectra at the relevant observation time. The best-fit parameters for the interpolated spectrum at each time are presented in Table~\ref{tbl:bestfitintp}. Vertical lines with endcaps indicate a subset of observational errors which are included for further insight into the $\chi^2$ fit results.}
\label{fig:at2017gfo}
\end{figure}

We identify the best-fitting parameters at each observation time $t$ using a simple $\chi^2$ goodness-of-fit statistic defined as 

\begin{equation}
\label{eq:chi2}
\chi^2 = \sum_{k = 0}^{1023} \left(\frac{F_{\lambda, \text{intp}, k} - \hat{F}_{\lambda, \text{obs}, k}}{\sigma_{\hat{F}_{\lambda, \text{obs}, k}}}\right)^2,
\end{equation}
where $k$ represents the \texttt{SuperNu} wavelength bins, $F_{\lambda, \text{intp}, k}$ is the interpolated spectral energy density scaled to 40 Mpc, $\hat{F}_{\lambda, \text{obs}, k}$ is the rebinned observed spectral energy density, and $\sigma_{\hat{F}_{\lambda, \text{obs}, k}}$ is the uncertainty on the observed spectral energy density. To assess the relative distribution of different model parameters $\vec{x}$, we use a likelihood $\exp(-\chi^2/2)$ and a uniform prior over ejecta parameters $\vec{x}$.  The samples $\vec{x}$ are iteratively drawn using Monte Carlo sampling (e.g., \cite{2023PhRvD.107b4040W}), and models are evaluated and compared to all wavelengths at each observation epoch.  From our posterior-weighted Monte Carlo samples, we use a maximum-likelihood estimate as the preferred value for $\vec{x}$, with statistical error bars on each component derived from the posterior distribution.

\begin{table}[ht]
\begin{center}
\begin{tabular}{cccccc}
\hline
$t$ & $\log_{10} M_d$ & $v_d$ & $\log_{10} M_w$ & $v_w$ & $\chi^2$/$N_t$ \\
{[days]} & [$M_\odot$] & [$c$] & [$M_\odot$] & [$c$] & \\ [0.5ex]
\hline\hline
\textbf{1.43} & $-1.47^{+0.11}_{-0.22}$ & $0.20^{+0.00}_{-0.00}$ & $-2.04^{+0.12}_{-0.00}$ & $0.10^{+0.01}_{-0.01}$ & 8538 \\
2.42 & $-2.05^{+0.00}_{-0.01}$ & $0.15^{+0.00}_{-0.00}$ & $-1.98^{+0.07}_{-0.12}$ & $0.18^{+0.00}_{-0.00}$ & 904 \\
3.41 & $-2.06^{+0.02}_{-0.03}$ & $0.19^{+0.10}_{-0.01}$ & $-1.91^{+0.03}_{-0.13}$ & $0.05^{+0.04}_{-0.00}$ & 539 \\
\textbf{4.4} & $-1.52^{+0.00}_{-0.00}$ & $0.11^{+0.00}_{-0.00}$ & $-1.51^{+0.00}_{-0.00}$ & $0.21^{+0.00}_{-0.00}$ & 957 \\
5.4 & $-1.71^{+0.00}_{-0.00}$ & $0.25^{+0.00}_{-0.00}$ & $-1.80^{+0.00}_{-0.00}$ & $0.09^{+0.00}_{-0.00}$ & 389 \\
6.4 & $-1.73^{+0.03}_{-0.00}$ & $0.14^{+0.01}_{-0.01}$ & $-1.81^{+0.00}_{-0.00}$ & $0.05^{+0.00}_{-0.00}$ & 238 \\
\textbf{7.4} & $-1.61^{+0.07}_{-0.04}$ & $0.29^{+0.00}_{-0.01}$ & $-1.80^{+0.01}_{-0.00}$ & $0.06^{+0.00}_{-0.01}$ & 385 \\
8.4 & $-2.05^{+0.11}_{-0.05}$ & $0.07^{+0.02}_{-0.00}$ & $-1.57^{+0.00}_{-0.01}$ & $0.09^{+0.00}_{-0.00}$ & 137 \\
9.4 & $-1.47^{+0.01}_{-0.04}$ & $0.30^{+0.00}_{-0.01}$ & $-1.80^{+0.01}_{-0.00}$ & $0.25^{+0.00}_{-0.00}$ & 155 \\
\textbf{10.4} & $-1.32^{+0.01}_{-0.00}$ & $0.30^{+0.00}_{-0.00}$ & $-2.05^{+0.07}_{-0.06}$ & $0.21^{+0.01}_{-0.00}$ & 45 \\ \hline 
\end{tabular}
\end{center}
\label{tbl:bestfitintp}
\caption{Best-fit parameters, \edited{with 1-$\sigma$ uncertainties}, derived from the comparison of interpolated spectra $F_{\lambda, \rm intp}$ to each of the ten X-shooter observational spectra $\hat{F}_{\lambda, \rm obs}$. Each set of parameters was separately identified and compared only to the spectrum taken at the observation time. Entries in bold have their spectra plotted in Figure~\ref{fig:at2017gfo}. \edited{All fits to spectra assume only a two-component model \emph{without} the inclusion of the additive thermal component}.}
\end{table}

\begin{figure}
\centering
\includegraphics[width=\columnwidth]{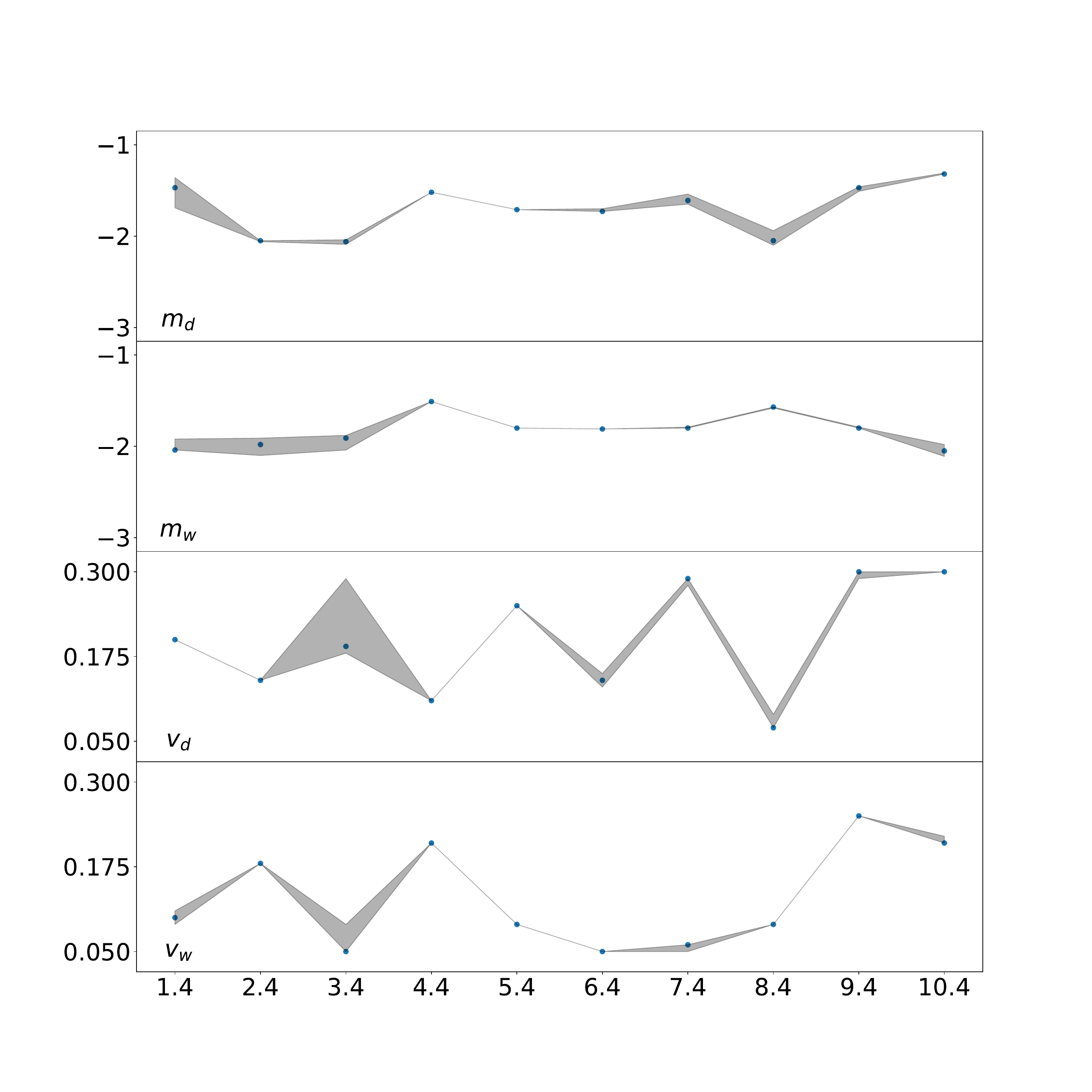}
\caption{\edited{Visual representation of the best-fit recovered parameters and their uncertainties presented in Table~\ref{tbl:bestfitintp}. The masses are fairly consistent across observation epochs, with wind mass slightly more stable than dynamical mass. Velocities are highly variable across observation epochs and can generally be considered poorly constrained. However, the wind velocity shows some consistency between 5-8 days, with a similar pattern seen in the wind mass at these times.}}
\label{fig:uncertainties}
\end{figure}

\begin{figure}
\centering
\includegraphics[width=\columnwidth]{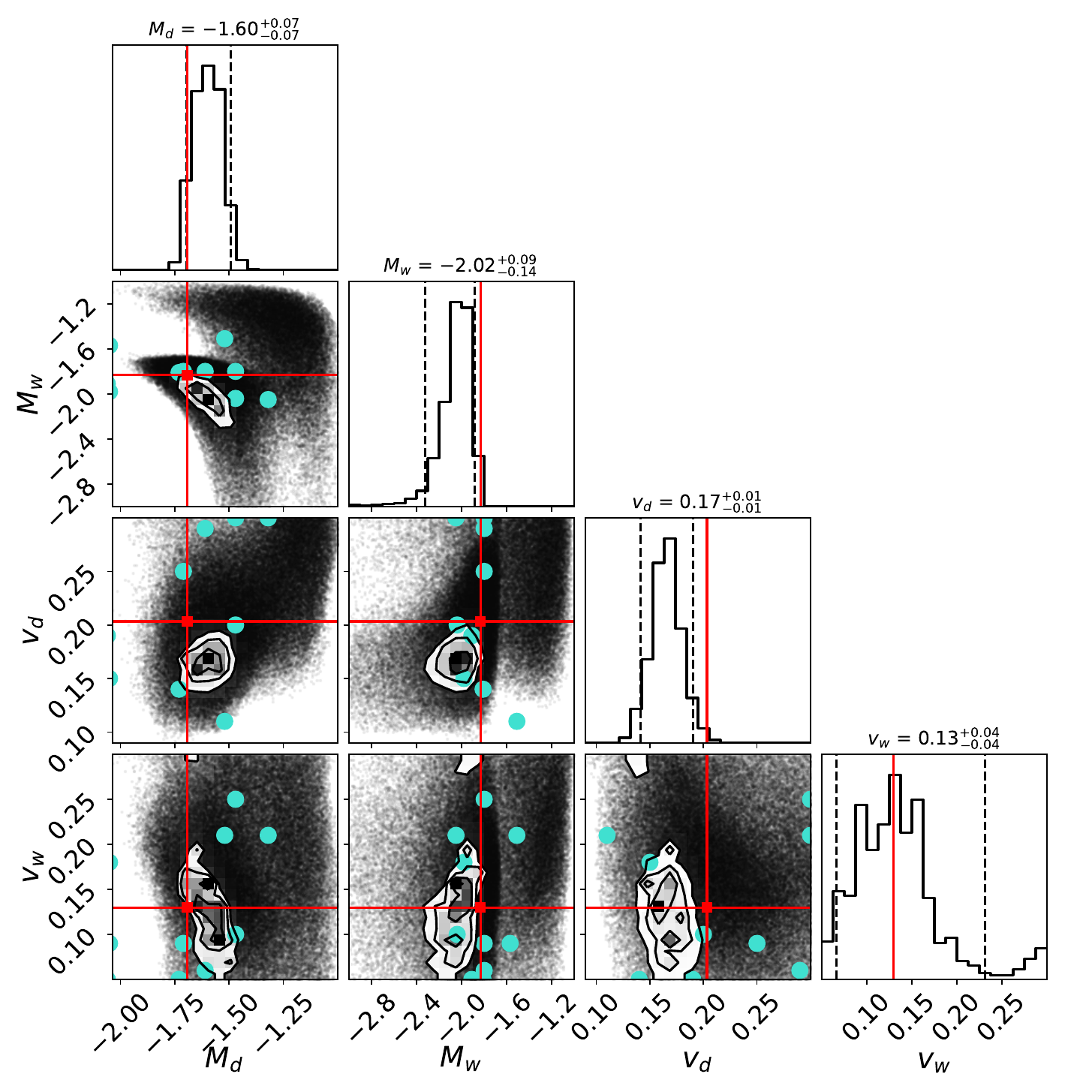}
\caption{Corner plot showing parameter recovery results from \cite{Ristic22} when omitting the K-band. The parameter means reported at the top of each parameter column represent the posterior distributions and their 90\% confidence intervals. Overlaid in red are weighted-average parameters calculated from the per-observation recovered parameters presented in Table~\ref{tbl:bestfitintp}.}
\label{fig:corner}
\end{figure}

Our two-component model fits to the AT2017gfo observational data are presented in Figure~\ref{fig:at2017gfo}. Early-time fits match well, especially at $1.43$ days where the spectrum effectively behaves like a blackbody. A notable discrepancy in the fit occurs at $1.43$ days in the $g$-band where our simulations are slightly underluminous around $0.4$~microns. At later times, this discrepancy becomes more exaggerated as the fit is increasingly underluminous in the $g$- and even $r$-bands at $7.4$ days. However, as time increases, our models nominally fit the data better, simply because of the relatively large measurement uncertainties at late times.  This nominally better statistical fit should not be taken as necessarily a more reliable parameter estimate, as for example at late times the local thermodynamic equilibrium assumption for our simulations becomes less applicable.

In Table~\ref{tbl:bestfitintp} we present the best-fitting model parameters, calculated using Equation~\ref{eq:chi2}, for the observed spectrum $\hat{F}_{\lambda, \rm obs}$ \edited{(labeled $F_{\lambda, \rm AT2017gfo}$ in the plot legend)} at each respective time. We also present the recovered parameters along with their uncertainties visually in Figure~\ref{fig:uncertainties} for clearer understanding of the parameter recovery differences at individual times. The $\chi^2/N_t$ values come directly from Equation~\ref{eq:chi2}; $N_t$ is a normalizing factor representing the number of wavelength bins used for comparison (up to 361) for the observation at time $t$. The $N_t$ normalizing factor accounts for the variable number of wavelength bins considered during the residual calculation for each observation time. The $\chi^2/N_t$ values shown in Table~\ref{tbl:bestfitintp} quantify the poor fit between data and our models seen in Figure~\ref{fig:at2017gfo} and elsewhere.  These large scaled residuals reflect the small observational uncertainties, as shown in Figure~\ref{fig:at2017gfo}, but as noted are also computed by completely neglecting any systematic error associated with either our interpolation or modeling. While we cannot thoroughly propagate our systematics at present, we estimate, based on small changes in our result to operating-point choices, as seen in Figure~\ref{fig:off_sample}, that incorporation of systematic error could account for much of the variation between our models and the data apparent at most wavelengths longer than 0.5 microns. The maximum systematic uncertainty for wavelengths less than 0.5 microns is $\Delta F_\lambda \sim 10^{-20}$, calculated as the maximum difference between predictions for the two models presented in Figure~\ref{fig:off_sample}. Therefore, we are confident that the underluminosity in the blue bands is indeed real and not simply due to modeling uncertainty. Decreasing $\chi^2/N_t$ at later times also not necessarily indicate better agreement between predictions and observations, but rather larger observational errors as spectra get increasingly noisier at these times. The \edited{non-uniformity} of the \edited{recovered} parameters \edited{is due to} each set of parameters \edited{being} identified at \edited{its relevant} observation time without regard to information from other \edited{times}. As such, it is difficult to make any explicit claims; however some trends do arise.

In particular, the dynamical mass tends to be greater than the wind mass for \edited{approximately half of the spectra.} \edited{The wind mass is the most consistent across observation epochs. We interpret our less variable constraints on wind mass as reflecting the wind ejecta radiation being prominent at earlier times where our fits to the spectra are better. Due to high opacity in the region, dynamical ejecta photons are expected to be emitted at later times; however, the data and our fit quality degrade at these times, leaving the dynamical ejecta properties more prone to variation compared to those of the wind ejecta. Velocities are overall highly variable across observations.}

\edited{To determine an aggregate set of ejecta parameters informed by inference at all observational times}, we calculate an overall residual from all spectra weighted by the number of points $N_t$ in each fit. We report weighted-average parameters $x$ such that $x = \sum_t N_t x_t / \sum_t N_t$, where each parameter $x$ is determined by the weighted sum of the recovered parameter at each time $x_t$, with $N_t$ serving as the weighting factor. The averaged parameters are presented in Figure~\ref{fig:corner}, overlaid on top of parameter recovery posteriors from the \cite{Ristic22} analysis, which excludes the $K$-band. \edited{The average parameters with uncertainties at the top of each posterior correspond to the \cite{Ristic22} results.} We find similar agreement for recovered parameters between the two analyses, \edited{with the understanding that the overlaid parameters are subject to the uncertainties from Table~\ref{tbl:bestfitintp}}.

\section{Three-Component Analysis}
\label{sec:3c}

The blue-wavelength underluminosity displayed in Figure~\ref{fig:at2017gfo} confirms that our detailed self-consistent radiative transfer simulations underpredict
the shortest optical-wavelength radiation at late times, both spectroscopically and photometrically \citep{Ristic22, 2023PhRvR...5a3168K}. \edited{This underprediction serves as a clear indicator that our modeling approach is missing an energy source that will sustain blue emission to late times without affecting the rest of the spectrum.} \edited{With the hypothesis that our two-component model composition assumptions are currently insufficient, we consider a third radioactive heating component as a natural extension of our existing model. To guide our parameter choices for the third component, we consider the effects of adding the flux from the simple kilonova model presented in \cite{Metzger_2019} to our spectra.}

\subsection{Simple Model for Parameter Guidance}
\label{sec:3c_metzger}

\edited{The kilonova model by \cite{Metzger_2019}, hereafter referred to as M19, calculates the blackbody spectral energy density at some time $t$ given an ejecta mass $M_{\rm ej}$, velocity $v_{\rm ej}$, and opacity $\kappa_{\rm ej}$. In the context of our study, a low-opacity third component is most preferable as it increases the likelihood of emission of blue photons rather than scattering or absorption. Likewise, a slow-moving component ensures that the blue-photon emitting ejecta does not diffuse too quickly, allowing for sustained blue emission at late times. Finally, the mass parameter acts as a scale factor for the overall brightness of the blackbody's spectral energy density.} 

Based on our fits to the spectra at all times, a subset of which is presented in Figure~\ref{fig:at2017gfo}, we identify that a gray-opacity model with $\kappa = 1$ cm$^2$/g and ejecta parameters $M_{\rm ej} = 0.003 M_\odot$ and $v_{\rm ej} = 0.005$c produces enough flux in the $g$- and $r$-bands to remedy the underluminosity without boosting the longer-wavelength flux, which our models match well. The spectral energy density $F_{\lambda, \rm M19}$ emitted by this component is simply added to our best-fit spectra $F_{\lambda, \rm intp}$ as a post-processing step, ignoring any potential photon reprocessing effects which may occur during radiative transfer.

\begin{figure}
\centering
\includegraphics[page=1, width=.95\columnwidth]{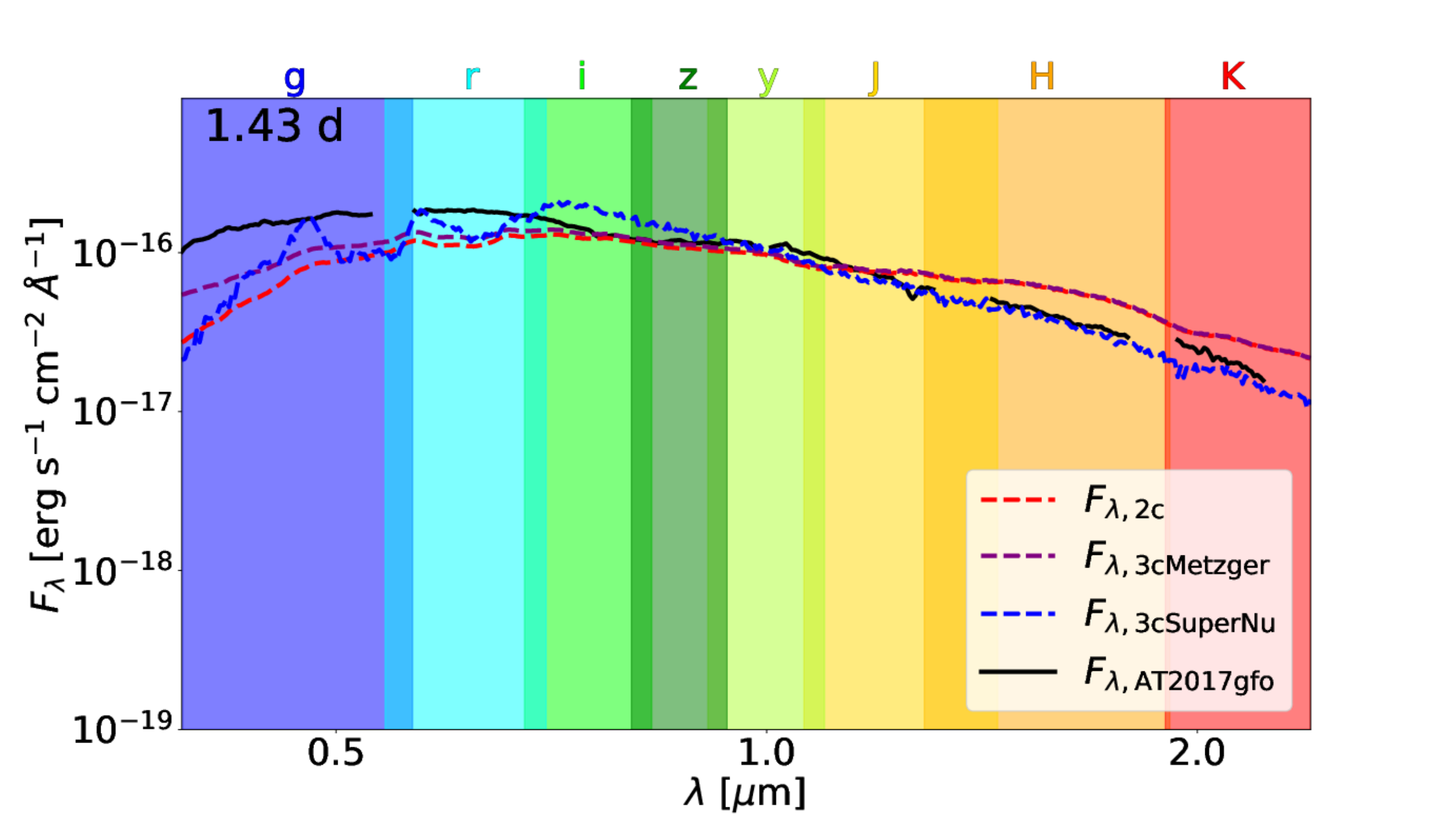}
\includegraphics[page=4, width=.95\columnwidth]{spec_intp_2c_vs_3c_optim_trimdata.pdf}
\includegraphics[page=7, width=.95\columnwidth]{spec_intp_2c_vs_3c_optim_trimdata.pdf}
\includegraphics[page=10, width=.95\columnwidth]{spec_intp_2c_vs_3c_optim_trimdata.pdf}
\caption{\edited{All spectral fits considered in this work. $F_{\lambda, \rm 2c}$ are the same two-component fits as in Figure~\ref{fig:at2017gfo}. The $F_{\lambda, \rm 3cMetzger}$ fits show the two-component fits with an additional third component flux contribution from the \cite{Metzger_2019} model with $M_{ej} = 0.003M_\odot$, $v_{ej} = 0.005$c, and $\kappa = 1$ cm$^2$/g. The $F_{\lambda, \rm 3cSuperNu}$ fit shows the \texttt{SuperNu} radiative transfer calculation of the M19 third component with closest-matching parameters $M_{ej} = 0.003$, $v_{ej} = 0.05$ and composition as shown in Figure~\ref{fig:eff_comps}.}}
\label{fig:at2017gfo_3c}
\end{figure}

Figure~\ref{fig:at2017gfo_3c} displays our best-fit interpolated spectra when including the additive thermal component from M19
during the residual calculation. The very-early and very-late spectra at $1.43$ and $10.4$ days exhibit little change with the addition of the third component in our relevant bands. The most obvious improvement occurs at $4.4$ days where the fit almost perfectly matches observations, but the $g$- and $r$-band underluminosity reappears in the $7.4$ day spectra. \edited{It is likely that the drop-off at $7.4$ days and later occurs due to the simplified approach of just adding the third component's spectral energy density to our existing best-fit spectra. }\edited{In order to understand the realistic, fully physical inclusion of the third component, we require a full radiative transfer calculation of our three-component model using \texttt{SuperNu}.}

\subsection{\texttt{SuperNu} Third Component}
\label{sec:3c_supernu}

The post-facto addition of a third component's flux contribution neglects important emission effects that can arise as a result of photon reprocessing in the ejecta. To consider the full physicality of including a third component, we present a \texttt{SuperNu} simulation involving a three-component model.

\begin{figure}
\centering
\includegraphics[width=\columnwidth]{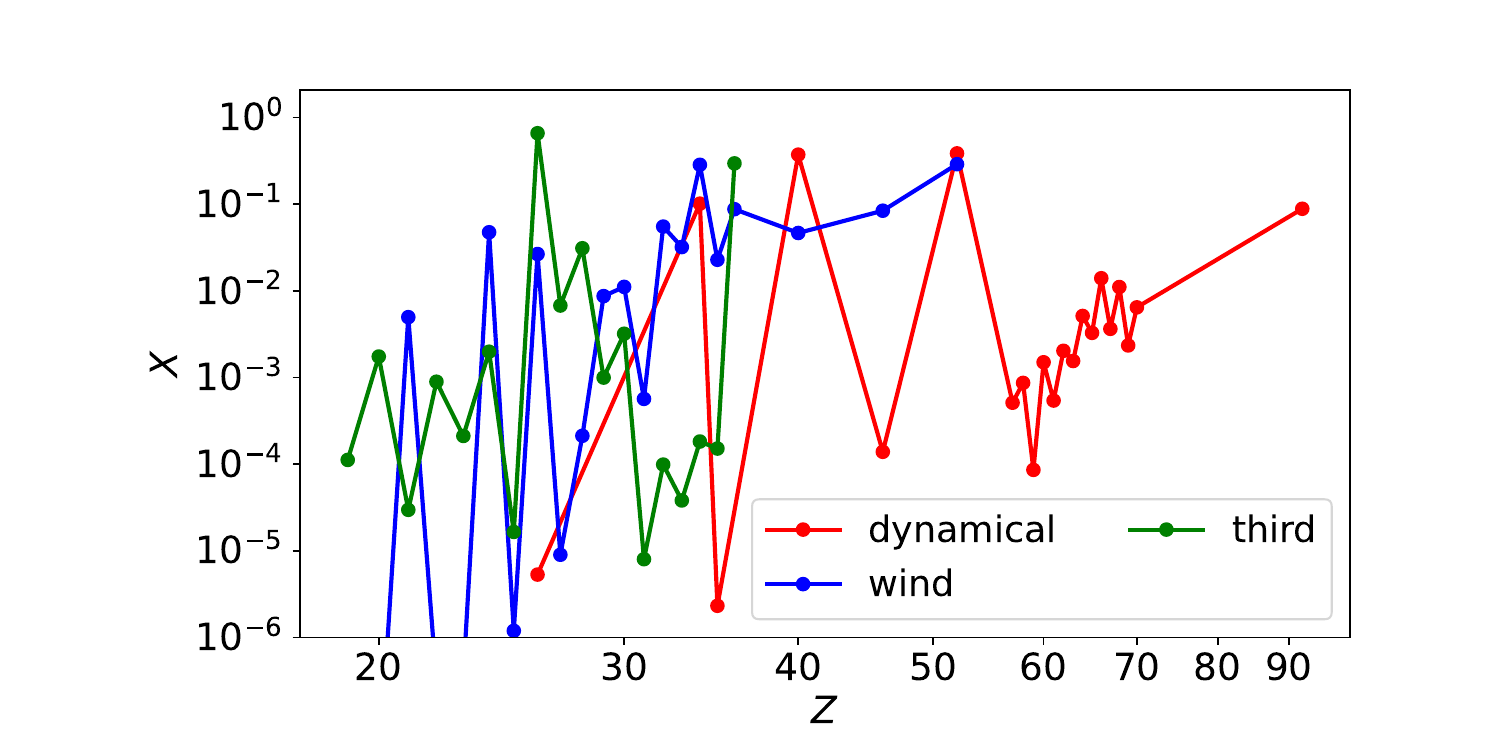}
\caption{Mass fractions $X$ as a function of element number $Z$ for the dynamical, wind, and third component compositions as described in Section~\ref{sec:3c_supernu}. The primary contribution of the third component comes from the large amount of iron ($Z=26$) and nickel ($Z=28$) which are not as prevalent in the other two components.}
\label{fig:eff_comps}
\end{figure}

Our three-component  \texttt{SuperNu} setup is an extension of our two-component approach. Our dynamical and wind component compositions remain unchanged and retain the properties described in Section~\ref{sec:sim_setup}. We incorporate the third component by mixing it into the  dynamical and wind components. For the third component, rather than considering a simple gray opacity as in the toy model, we use the detailed line-binned opacities described in Section~\ref{sec:sim_setup}, associated with a low-opacity, lanthanide-free composition shown by the green line in Figure~\ref{fig:eff_comps}. Due to the similarity between the dynamical and wind ejecta heating rates, we employ the dynamical ejecta heating rate to both the dynamical and wind components for computational simplicity. The composition and heating rate for the third component were generated using the \texttt{WinNet} nuclear reaction network for a homologously expanding ejecta with a velocity of $0.05c$ and characterized by electron fraction $Y_e = 0.50$.

The averaged, aggregate parameters for the \edited{dynamical and wind components for the original two-component model} are taken from Figure~\ref{fig:corner}.
The mass of the third component is fixed to $M_{\rm ej} = 0.003$ as in Section~\ref{sec:3c_metzger}. The third component velocity $v_{\rm ej}$ is increased to $0.05c$ to match the lowest allowed value in the \texttt{SuperNu} velocity space. Increasing $v_{\rm ej}$ from 0.005 to 0.05 also prevents ejecta fallback onto the remnant. As discussed in \cite{2019ApJ...880...22W}, ejecta fallback would require an additional energy source treatment and remove our assumption of a single radioactive-heating energy source. 

Figure~\ref{fig:at2017gfo_3c} shows all of the different spectra modeling efforts considered in this work compared to the AT2017gfo observed spectra. The ``2c" spectra match the two-components fits presented in Figure~\ref{fig:at2017gfo}, the ``3cMetzger" spectra are the best-fit ``2c" spectra, which include the additive thermal component from M19, and the ``3cSuperNu" spectra present fits from the \texttt{SuperNu} run that uses the third component described in the preceding paragraph. Starting as early as $4.4$ days, it is obvious that the self-consistent implementation of the third component in \texttt{SuperNu} does not provide nearly as much short-wavelength flux as the Metzger additive thermal component.

In fact, for the majority of observation times, the ``3cSuperNu" model is even \emph{less luminous} than the ``2c" model, instead shifting spectral energy density from blue wavelengths to redder ones. \edited{This shift seems to indicate that the inclusion of the third component is reprocessing photons to longer wavelengths instead of amplifying the emission at shorter ones}. At $10.4$~days, the massive spike in flux at $1.5$~microns also indicates that our third component is not optimally suited to matching the features of the AT2017gfo spectra. 

\edited{Given the results of Figure~\ref{fig:at2017gfo_3c}, we find that an additional radioactive component is not sufficient to amplify, or even match, the required flux to match our models to the AT2017gfo data. The reprocessing of photons to lower energies in the additional component introduces an unwanted flux boost around $1.5$~microns, which results in even worse-fitting spectra than those using only two components. As such, future studies should explore detailed composition analysis to achieve an increase in blue emission within the constraints of the two-component model. Likewise, Figure~\ref{fig:at2017gfo_3c} is an illustrative example that an additional modeling component may not necessarily be a radioactive heating source. 

A notable caveat is that our third component was initially chosen to have a slow velocity in order to boost late-time blue emission; a similar radioactive-heating component with a velocity faster than that of the wind ejecta may exhibit fewer photon reprocessing effects to longer wavelengths by virtue of the photons not having to interact with the wind component as they escape.}

\section{Conclusions}
\label{sec:conclusion}

We have demonstrated that a straightforward approach can accurately interpolate between simulated spectra derived from
radiative-transfer simulations of kilonova
ejecta across a high-dimensional model parameter space.  In this proof-of-concept study, motivated by the relative
scarcity of spectral observations, we fix the spectra viewing angle (time) and only interpolate over ejecta properties spanning four dimensions and time (angle) spanning one dimension, \edited{applicable in both scenarios given our assumption of axisymmetry}. 

Although this work focused specifically on kilonova spectra, the interpolation scheme should be broadly applicable to all astrophysical spectra of similar dimensionality. While our initial  highly non-parsimonious approach produces accurate spectra, we find that its large memory footprint and computational cost can be substantially reduced. The nature of the large dataset would make it well-suited for conventional machine-learning techniques, such as neural networks.

We have used our interpolated spectra to recover the closest-matching model parameters \edited{that} replicate the observed spectra of kilonova AT2017gfo. We present multiple modeling approaches, including a standard two-component approach, a three-component approach using an additive third component, and a three-component approach implemented in the Monte Carlo radiative transfer code \texttt{SuperNu}. In accordance with our previous parameter inference study \citep{Ristic22}, as well as other studies of a similar nature \citep{2013ApJ...776L..40Y, 2014MNRAS.439.3916M, 2018ApJ...861L..12L, 2018ApJ...861...55M, 2019MNRAS.483.1912P, 2019ApJ...880...22W, 2021ApJ...915L..11A, Nicholl21}, we find that an additional modeling component is necessary to overcome early-time underluminosity in the $g$- and $r$-bands. With the inclusion of the relatively light, slow-moving, lanthanide-free component, the short-wavelength spectral energy distribution remains underluminous at later times, with a clear discrepancy already present at a week post-merger. The \edited{persistent} $g$- and $r$-band disagreement at late times implies \edited{that} \edited{an additional radioactive component is not a suitable modeling approach, indicating the need for} a more sophisticated treatment of the blue-wavelength flux contribution in further studies.

Finally, in this paper, our analysis highlights future studies which will expand our composition assumptions in order to better understand the impact of ejecta composition on the blue flux contribution. However, there are many other uncertainties associated with the models, such as mass and composition distributions as a function of velocity and angle, atomic physics results assuming local thermodynamic equilibrium, and the finer treatment of energy deposition into the ejecta via different decay channels. As we learn about new sensitivities from these uncertainties, it becomes increasingly clear that it will be difficult to create a fine grid of models covering all of these effects. Our method is useful for the applications outline in this paper, but also because it can ultimately be scaled to adapt to the wider parameter space of model uncertainties, using a limited number of simulations to intelligently map between results. 

\section{Acknowledgments}
ROS and MR acknowledge support from NSF AST 1909534. ROS acknowledges support from NSF AST 2206321. VAV acknowledges support by the NSF through grant AST-2108676. The work by CLF, CJF, OK, and RTW was supported by the US Department of Energy through the Los Alamos National Laboratory (LANL). This research used resources provided by LANL through the institutional computing program. Los Alamos National Laboratory is operated by Triad National Security, LLC, for the National Nuclear Security Administration of U.S.\ Department of Energy (Contract No.\ 89233218CNA000001).

\bibliographystyle{aasjournal}
\bibliography{bibliography,thesis,LIGO-publications,gw-astronomy-mergers-ns-gw170817}

\end{document}